\documentclass[lettersize,journal]{IEEEtran}
\usepackage{amsmath,amsfonts}
\usepackage{algorithmic}
\usepackage{algorithm}
\usepackage{amsthm}
\usepackage{amssymb}
\usepackage{booktabs}
\usepackage{xcolor}
\usepackage{multirow} 
\usepackage{algorithm}

\usepackage{booktabs}
\usepackage{multirow}
\usepackage{graphicx}

\theoremstyle{definition}

\theoremstyle{theorem}

\theoremstyle{lemma}

\theoremstyle{Assumption}

\DeclareMathOperator{\Pool}{Pool}
\DeclareMathOperator{\LN}{LN}

\usepackage{array}
\usepackage[caption=false,font=normalsize,labelfont=sf,textfont=sf]{subfig}
\usepackage{textcomp}
\usepackage{stfloats}
\usepackage{url}
\usepackage{verbatim}
\usepackage{graphicx}
\usepackage{cite}
\usepackage{color}
\usepackage{soul}

\begin{document}
\title{Multi-Source and Multi-Dimensional Information Fusion in Resource-Constrained Edge Networks}


\author{Yushi Wang, Zhengxin Yu,~\IEEEmembership{Member,~IEEE}, Guhan Zheng,~\IEEEmembership{Member,~IEEE}, Wang Miao,~\IEEEmembership{Member,~IEEE}, \\Merouane Debbah,~\IEEEmembership{Fellow,~IEEE}

\thanks{Yushi Wang and Zhengxin Yu are with the School of Computing and Communications, Lancaster University, LA1 4YW, U.K. (e-mail: y.wang216@lancaster.ac.uk; z.yu8@lancaster.ac.uk).}
\thanks{Guhan Zheng is with the School of Communication and Information Engineering, Shanghai University, 200444, China. (e-mail: gzheng@shu.edu.cn).}
\thanks{Wang Miao is with the Department of Computer Science, University of Exeter, EX4 4QF, U.K. (e-mail: wang.miao@exeter.ac.uk).}
\thanks{Merouane Debbah is with KU 6G Research Center, Khalifa University of Science and Technology, 127788, Abu Dhabi, UAE. (e-mail: merouane.debbah@ku.ac.ae).}}

\markboth{}%
{}
\maketitle

\begin{abstract}
Multi-source semantic information fusion enables edge networks to support global inference using observations collected by distributed terminals. However, limited communication and computation resources restrict semantic uploading, leaving the edge server with incomplete tokens that differ in modality, semantic granularity, arrival time, and channel condition. The value of these tokens also depends on the complementary or redundant information provided by other terminals. In this paper, we propose a multi-source and multi-dimensional semantic information fusion framework that coordinates terminal uploading and edge-side fusion under resource constraints. We develop Hyper-Slot to combine semantic content with available token attributes through iterative token-to-slot assignment and slot-to-token feedback, producing a compact representation for global inference. To encourage useful semantic uploading, we further design a contribution-based incentive mechanism that uses Hyper-Slot assignments to evaluate information coverage and cross-type complementarity. A bi-level Stackelberg–Nash game determines the server reward pool and terminal transmission-quality strategies while accounting for computation and communication energy costs. Experiments on a tri-modal extension of COCO2014 and image-text retrieval on Flickr30K show that Hyper-Slot improves mAP by 3.66 percentage points and RSum by 15.9 points over the strongest evaluated baselines, respectively, while reducing the number of parameters by more than 80\% compared with a naive concatenation-based fusion baseline.



\end{abstract}


\section{Introduction}

Sixth-generation (6G) communication networks will support numerous intelligent applications, including intelligent transportation and vehicular networks \cite{10416926} and the unmanned aerial vehicles systems \cite{10445211}. These applications involve distributed terminals continuously generating various types of information, such as images, text, audio, and multisensor data  \cite{10431795}. Although the transmission of such information enables high service quality and new intelligent capabilities, it also significantly increases data transmission loads on edge networks with constrained spectrum resources. Efficiently utilizing information generated by distributed terminals therefore becomes an important issue for 6G networks.

Multi-source information fusion provides an effective approach to this issue by integrating information from different terminals into a unified representation for global inference \cite{xie2022task}. In edge networks, terminals can use local encoders to represent their observations as task-relevant semantic tokens, which are then transmitted to an edge server for fusion. Compared with relying on the information of a single terminal, multi-source fusion enables the global task to exploit information collected from different locations, sensing ranges, and data modalities. It is therefore a promising method for supporting intelligent tasks in 6G-enabled edge networks.

Directly applying conventional multi-source information fusion methods to resource and communication constrained edge networks presents several challenges. Existing methods usually assume that information from all terminals can be completely transmitted to the edge server \cite{khaleghi2013multisensor}. However, terminals in practical edge networks have limited computation capacity, transmission power, and available bandwidth. The edge server therefore cannot always obtain complete and high-quality semantic information from every terminal. The number and quality of the tokens available for fusion therefore depend on the uploading strategies of different terminals.

This problem is further complicated by the differences among terminals. Semantic information collected by different terminals may provide different contributions to the target task. Their channel conditions and resource costs may also vary significantly, so each terminal must determine whether to upload each token and at what transmission quality. Independent decisions may cause terminals to transmit redundant information while omitting important semantic evidence from other sources. Both cases reduce the coverage and complementarity of the information available at the edge server. It is therefore necessary to coordinate the upload decisions of multiple terminals and preserve valuable and complementary semantic information under limited resources.

Another challenge arises from the incomplete and heterogeneous semantic tokens produced by selective uploading. In addition to their semantic content, tokens may contain different attributes related to the fusion process. These attributes may include source, modality, semantic granularity, temporal state, and channel condition. The relevant dimensions vary with the application scenario, and together with the semantic content they shape each token's contribution to the task. For example, a semantically relevant token may become less useful under poor channel conditions, while several tokens with limited individual contributions may together provide important evidence. Conventional concatenation, pooling, and attention methods struggle to capture these relationships, and become less reliable when the number, quality, and structure of the received tokens vary across terminals.


These challenges are closely coupled. Terminal uploading decisions determine the information available at the edge server, while the value of uploading a token depends on how it is fused with information from other terminals. Therefore, an effective fusion framework for edge networks should jointly address selective semantic uploading and multi-dimensional information fusion. It should coordinate uploading under communication and computation constraints and exploit both semantic content and available token attributes to fuse incomplete and heterogeneous tokens into a task-level representation.

To address the above challenges, we propose a multi-source and multi-dimensional semantic information fusion framework for resource and communication constrained edge networks. The framework coordinates semantic uploading among multiple terminals and fuses the incomplete and heterogeneous tokens received at the edge server. The main contributions are summarized as follows:

\begin{itemize}

\item We propose a unified framework for multi-source and multi-dimensional semantic information fusion in resource-constrained edge networks. The framework jointly considers selective semantic uploading at distributed terminals and information fusion at the edge server. By using the fusion assignments to evaluate terminal contributions and guide uploading decisions, the framework connects selective semantic uploading with edge-side information fusion under communication and computation constraints.


\item We develop Hyper-Slot to aggregate incomplete and variable-size sets of heterogeneous semantic tokens into a fixed number of latent slots. Hyper-Slot jointly models semantic content and available token attributes to determine token-to-slot assignments. Through iterative token-to-slot assignment and slot-to-token feedback, it refines the grouping and aggregation of heterogeneous tokens to produce a compact task-level representation for global inference.


\item We design a semantic uploading incentive mechanism to encourage terminals to provide valuable and complementary information for global inference. The mechanism evaluates terminal contributions through the Hyper-Slot assignments, jointly considering information coverage, cross-type complementarity, and computation and communication energy costs. A bi-level Stackelberg–Nash game is then formulated to determine the server reward pool and the transmission-quality strategies of terminals under resource constraints.

\end{itemize}

\section{Related Works}

Edge intelligence enables terminals and edge servers to perform collaborative inference close to data sources. Zhou et al. \cite{zhou2019edge} reviewed the deployment of artificial intelligence services at the network edge. Shi et al. \cite{shi2020communication} further discussed the communication cost introduced by device-edge collaboration. To reduce this cost, Xie et al. \cite{xie2022task} developed a task-oriented multi-user semantic communication framework for image retrieval, machine translation, and visual question answering. Shao et al. \cite{shao2022task} studied multi-device cooperative inference through compact local features. More recent studies have moved from transmitting complete feature maps toward finer-grained semantic representations. CodeFilling \cite{hu2024communication} represents selected collaborative features using compact codebook indices, while CoSDH \cite{xu2025cosdh} selects collaboration regions according to the information supply and demand of different agents. These methods improve the trade-off between perception accuracy and communication overhead. However, they generally operate under predefined collaboration protocols and fixed communication budgets, where terminal participation, upload quality, and heterogeneous resource costs are taken as given rather than jointly optimized.

Resource-constrained semantic communication studies how semantic information can be efficiently delivered under limited communication and computation resources. DeepSC \cite{xie2021deep} and L-DeepSC \cite{xie2020lite} learn semantic representations that are robust to wireless channels. Deep joint source-channel coding has also been investigated for wireless image transmission \cite{bourtsoulatze2019deep}, while Jankowski et al. \cite{jankowski2020wireless} optimized transmitted representations for edge image retrieval. Recent studies further couple semantic importance with wireless conditions: Wang et al. \cite{wang2024feature} select task-important features and allocate channels to improve semantic spectral efficiency, and Yan et al. \cite{yan2024qoe} jointly optimize semantic compression, channel assignment, and transmit power for multi-task networks. These studies improve how efficiently and reliably semantic information is delivered under constrained resources. Their selection and allocation decisions, however, are driven by communication-side objectives such as spectral efficiency, transmit power, or latency, and remain decoupled from how the delivered information is fused at the receiver. The received tokens are typically handed to a predefined decoder or task module, so whether a transmitted token is worth its cost for the subsequent multi-source fusion is left unmodeled.

After transmission, the received semantic tokens must be integrated for downstream inference. Lee et al. \cite{lee2018stacked} used stacked cross-attention for image-text matching, while Fu et al. \cite{fu2023learning} modeled semantic relations at both fragment and instance levels. Set Transformer \cite{lee2019set} provides an attention-based architecture for set aggregation, and recent collaborative perception methods such as CoST \cite{tang2025cost} jointly aggregate information across agents and time steps in a unified spatiotemporal space. A separate line of work models structure explicitly: hypergraph neural networks \cite{feng2019hypergraph} aggregate features over predefined high-order connections among elements. Our fusion mechanism instead follows the slot-based, object-centric tradition, where a set of slots competitively aggregate input elements rather than relying on a fixed higher-order graph. Slot Attention \cite{locatello2020object} introduced this paradigm to discover latent groups from a single image, without communication, heterogeneous attributes, or incomplete inputs. These methods are each designed for a particular relation, such as cross-modal interaction, set aggregation, high-order structure, latent grouping, or spatiotemporal dependency. They do not directly provide a unified mechanism for combining semantic content with the auxiliary attributes available in different edge scenarios, which may describe the modality, semantic granularity, temporal status, or channel condition of a token. This motivates a flexible multi-dimensional fusion method that can model their joint influence on information quality and cross-source complementarity.





\begin{figure*}[t]
    \centering
        \centering
        \includegraphics[width=\textwidth]{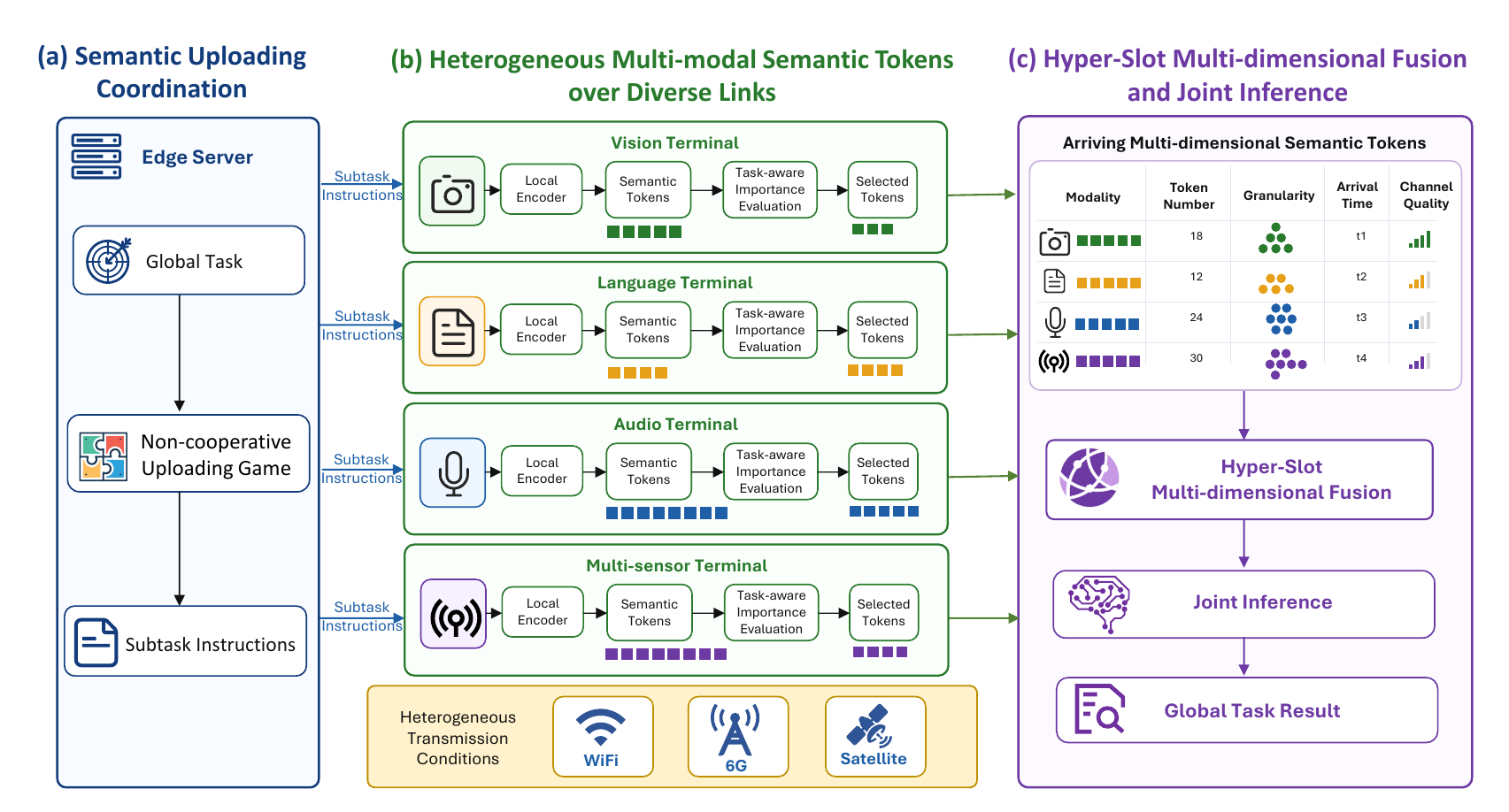}
    \caption{Overview of the proposed multi-source and multi-dimensional semantic information fusion framework.}
    \label{fig:framework}
    \centering
\end{figure*}

\begin{figure*}[t]
    \centering
        \centering
        \includegraphics[width=\textwidth]{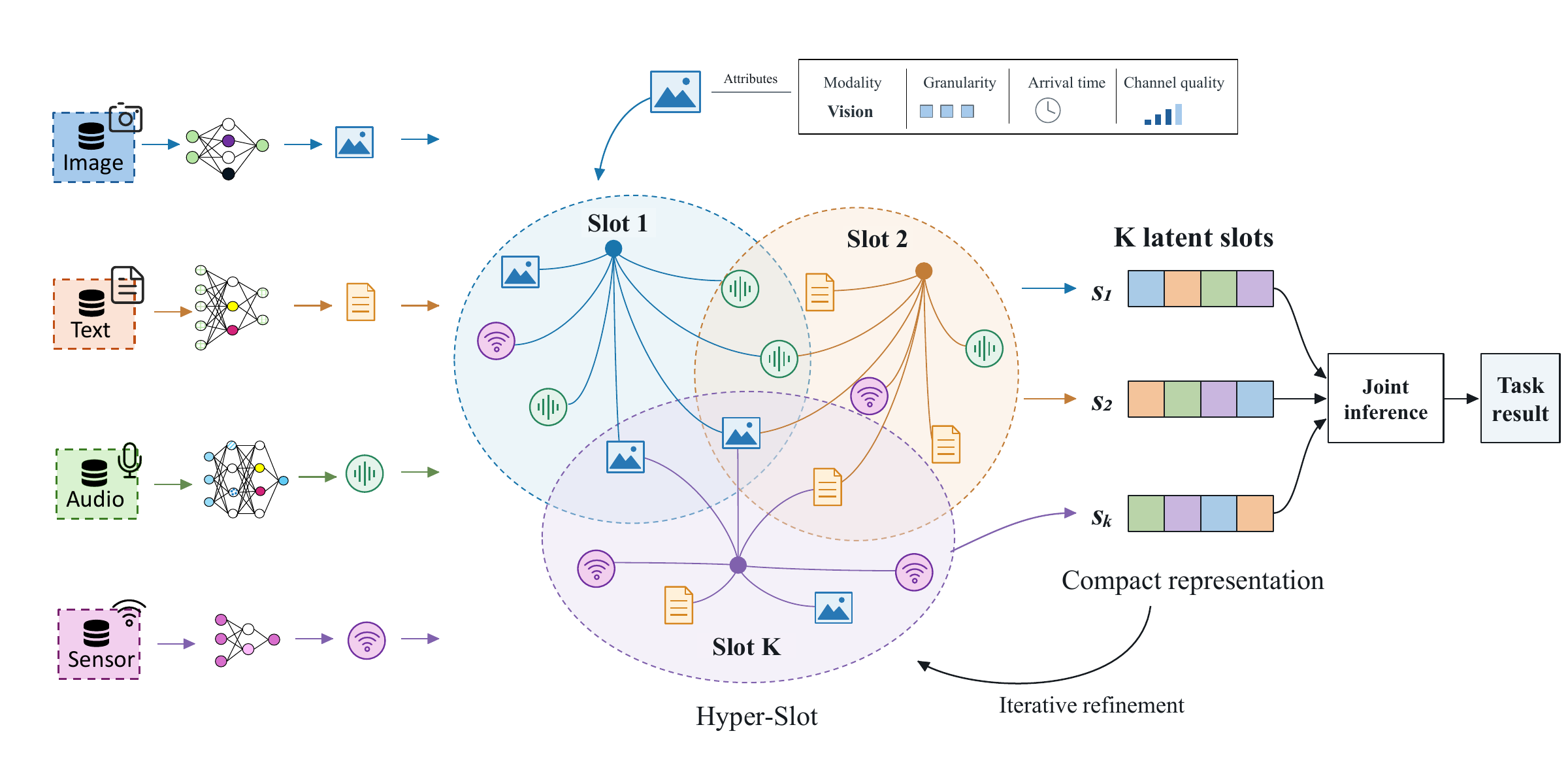}
    \caption{Structure of the proposed Hyper-Slot module. Semantic tokens are assigned to latent slots using semantic content and token attributes, and iteratively aggregated into a compact representation for global inference.}
    \label{fig:hyperslot}
    \centering
\end{figure*}

\section{System Model}
In this section, we describe the considered edge network and its multi-source semantic communication and fusion model. We then introduce the computation and communication overhead associated with semantic uploading.

\subsection{Multi-Source Semantic Communication and Fusion Model}

We consider an edge network consisting of one edge server and
$M$ terminals, indexed by
$m\in\mathcal{M}=\{1,\ldots,M\}$.
The system operates over discrete time slots
$t=1,\ldots,T$.
At time slot $t$, terminal $m$ collects a local observation
$\mathcal{O}_m^t$ and employs its local encoder to extract a
set of candidate semantic tokens:
\begin{equation}
\mathcal{X}_m^t
=
\operatorname{Encoder}_m(\mathcal{O}_m^t)
=
\left\{
\mathbf{x}_{m,1}^t,\ldots,
\mathbf{x}_{m,N_m^t}^t
\right\},
\end{equation}
where $N_m^t$ denotes the number of candidate tokens generated
by terminal $m$. Different terminals may observe the same task
from different locations, sensing ranges, or modalities. Their
local encoders may also have different structures. We assume
that the extracted tokens are projected into a common semantic
space, where
$\mathbf{x}_{m,i}^t\in\mathbb{R}^{d_x}$.

Each token $\mathbf{x}_{m,i}^t$ is associated with an attribute
descriptor $\eta_{m,i}^t$ and has a size of $s_{m,i}^t$ bits.
The descriptor records the information available for semantic
fusion. It may include the source, modality, semantic
granularity, temporal state, and channel condition of the
token. These attributes depend on the terminal, observation,
and transmission environment. Therefore, tokens generated by
different terminals may differ in semantic content, attribute
configuration, and data size.

Due to limited communication and computation resources,
terminal $m$ cannot upload all candidate tokens. Let
\begin{equation}
b_{m,i}^{t}\in[0,1].
\end{equation}
Here, $(b_{m,i}^{t})$ denotes the normalized transmission quality assigned to token $x_{m,i}^{t}$. A value of $b_{m,i}^{t}=0$ indicates that the token is not uploaded, while a larger positive value indicates higher transmission quality. The value $b_{m,i}^{t}=1$ represents the highest transmission quality. The uploading strategy
of terminal $m$ is represented by
\begin{equation}
\mathbf{b}_m^t
=
\left[
b_{m,1}^t,\ldots,b_{m,N_m^t}^t
\right].
\end{equation}
Accordingly, the token set selected by terminal $m$ is
\begin{equation}
\mathcal{X}_{m,\mathrm{up}}^{t}
    =
    \left\{
    (x_{m,i}^{t},\eta_{m,i}^{t})
    \mid b_{m,i}^{t}>0,\ i=1,\ldots,N_m^{t}
    \right\}.
\end{equation}

We consider the uplink transmission from terminal $m$ to the
edge server over a wireless channel with bandwidth $B_m^t$,
transmit power $p_m^t$, and channel gain $h_m^t$. The
achievable uplink rate is
\begin{equation}
R_m^t
=
B_m^t
\log_2
\left(
1+
\frac{p_m^t h_m^t}{\sigma^2}
\right),
\end{equation}
where $\sigma^2$ is the noise power and
$0\leq p_m^t\leq P_m^{\max}$.
The channel condition can also be included in
$\eta_{m,i}^t$ as a communication-related attribute.

After transmission, the edge server receives the selected
tokens from all terminals. The received token set is
\begin{equation}
\begin{aligned}
    \mathcal{X}^t
&=
\bigcup_{m\in\mathcal{M}}
\mathcal{X}_{m,\mathrm{up}}^t
\\&=
\left\{
\left(
\mathbf{x}_{m,i}^t,\eta_{m,i}^t
\right)
\mid
b_{m,i}^{t}>0,\;
m\in\mathcal{M},\;
i=1,\ldots,N_m^t
\right\}.
\end{aligned}
\end{equation}
The size and composition of $\mathcal{X}^t$ depend on the
uploading strategies
$\{\mathbf{b}_m^t\}_{m\in\mathcal{M}}$.
Hence, the edge server generally receives an incomplete and
variable-size token set. The received tokens may provide
complementary information or describe overlapping semantic
content.

The edge server applies an edge-side fusion and inference
mapping to the received token set:
\begin{equation}
\widehat{\mathbf{y}}^t
=
\mathcal{G}
\left(
\mathcal{X}^t
\right),
\end{equation}
where $\mathcal{G}(\cdot)$ denotes the general mapping from
the received multi-source semantic tokens to the global task
result $\widehat{\mathbf{y}}^t$. Its output depends jointly on
the semantic content, available attributes, and uploading
decisions of all terminals. The specific fusion implementation
is introduced in the following section.

\subsection{Computation and Communication Overhead Model}

Terminal $m$ incurs computation overhead when processing
the selected semantic tokens and communication overhead when
uploading them to the edge server. Let $c_{m,i}^t$ denote the number of CPU cycles required to
process token $\mathbf{x}_{m,i}^t$ for uploading. This process
may include token refinement, attribute processing, and
transmission preparation. The total computation load of
terminal $m$ is
\begin{equation}
C_m^t
=
\sum_{i=1}^{N_m^t}
b_{m,i}^t c_{m,i}^t.
\end{equation}
Let $f_m^t$ denote the CPU frequency allocated by terminal
$m$. The corresponding computation delay is
\begin{equation}
\tau_{m,\mathrm{cmp}}^t
=
\frac{C_m^t}{f_m^t}.
\end{equation}
The computation energy consumption is given by 
\begin{equation}
e_{m,\mathrm{cmp}}^t
=
\kappa_m
\left(f_m^t\right)^2
C_m^t,
\end{equation}
where $\kappa_m$ is the effective switched-capacitance
coefficient of terminal $m$ \cite{mao2017survey}.

The total size of the semantic tokens uploaded by terminal
$m$ is
\begin{equation}
S_m^t
=
\sum_{i=1}^{N_m^t}
b_{m,i}^t s_{m,i}^t.
\end{equation}
Based on the achievable rate $R_m^t$, the communication
delay is
\begin{equation}
\tau_{m,\mathrm{com}}^t
=
\frac{S_m^t}{R_m^t}
=
\frac{
\sum_{i=1}^{N_m^t}
b_{m,i}^t s_{m,i}^t
}{
R_m^t
}.
\end{equation}
The corresponding communication energy consumption is
\begin{equation}
e_{m,\mathrm{com}}^t
=
p_m^t\tau_{m,\mathrm{com}}^t
=
\frac{
p_m^t
\sum_{i=1}^{N_m^t}
b_{m,i}^t s_{m,i}^t
}{
R_m^t
}.
\end{equation}

Therefore, the total processing delay and energy consumption
of terminal $m$ are respectively expressed as
\begin{equation}
\tau_m^t
=
\tau_{m,\mathrm{cmp}}^t
+
\tau_{m,\mathrm{com}}^t,
\end{equation}
and
\begin{equation}
e_m^t
=
e_{m,\mathrm{cmp}}^t
+
e_{m,\mathrm{com}}^t.
\end{equation}

\section{The Proposed Framework}
In this section, we present the proposed framework for multi-source and multi-dimensional semantic information fusion. We first describe the overall framework, followed by the Hyper-Slot fusion module and the semantic uploading mechanism based on a bi-level Stackelberg--Nash game.

\subsection{Framework Description}
\label{sec:framework}
Fig.~\ref{fig:framework} shows the proposed multi-source and multi-dimensional information fusion framework. At each time slot $t$, every terminal $m\in\mathcal{M}$ generates a candidate semantic token set $\mathcal{X}_m^t$ from its local observation, and selects which tokens to upload under its own communication and computation constraints.

The tokens received at the edge server are generally incomplete, variable in size, and heterogeneous in modality, semantic granularity, arrival time, and channel condition. To fuse such tokens, we propose Hyper-Slot in Section~\ref{sec:hyperslot}, which combines each token's semantic content with its available attributes and aggregates the received tokens into a
unified task-level representation, as shown in  Fig.~\ref{fig:hyperslot}. 


Since Hyper-Slot aggregates tokens according to their joint contribution to each slot, a terminal's effect on the fused representation depends on what the other terminals upload as well, so independent per-terminal uploading decisions can leave the received tokens redundant or incomplete. We therefore coordinate uploading through a two-level game in Section~\ref{sec:game}. The edge server announces an incentive, and the terminals determine their uploading strategies under this incentive through a non-cooperative game that settles at a Nash equilibrium. The reward given to each terminal is computed from its marginal contribution to the fusion gain, which is defined directly from the Hyper-Slot assignment weights. This ties each terminal's uploading decision to the information already provided by the others, allowing the framework to retain complementary information and reduce redundant transmission under resource constraints.

At each time slot, the terminals first generate candidate tokens, the
two-level game then determines their uploading strategies, and the selected tokens are fused by Hyper-Slot to yield the global inference result. The game determines which information reaches the fusion stage, while Hyper-Slot determines how that information is combined.

\begin{algorithm}[htbp]
\caption{The Proposed Framework}
\label{alg:framework}
\begin{algorithmic}[1]
\REQUIRE Local observations $\{O_m^t\}$, time slots $T$,
slot number $K$, refinement iterations $L$

\FOR{$t=1$ to $T$}
    \FORALL{$m\in\mathcal{M}$}
        \STATE $\mathcal{X}_m^t
        \leftarrow
        \mathrm{Encoder}_m(O_m^t)$
    \ENDFOR

    \STATE $\{\mathbf{b}_m^{t*}\}_{m\in\mathcal{M}}
    \leftarrow
    \mathrm{GAME}
    \bigl(\{\mathcal{X}_m^t\}_{m\in\mathcal{M}},\mathcal{G}\bigr)$

    \FORALL{$m\in\mathcal{M}$}
        \STATE $\mathcal{X}_{m,\mathrm{up}}^t
        \leftarrow
        \{(\mathbf{x}_{m,i}^t,\eta_{m,i}^t)
        \mid b_{m,i}^{t*}>0\}$
    \ENDFOR

    \STATE $\mathcal{X}^t
    \leftarrow
    \displaystyle\bigcup_{m\in\mathcal{M}}
    \mathcal{X}_{m,\mathrm{up}}^t$

    \STATE $\{(\mathbf{x}_i,\eta_i)\}_{i=1}^{N^t}
    \leftarrow
    \mathcal{X}^t$

\FORALL{$i=1,\ldots,N^t$}
        \STATE $\mathbf{d}_i
        \leftarrow
        \Gamma_{\theta}(\eta_i)$
        \STATE $g_i
        \leftarrow
        \mathrm{Gate}(\mathbf{d}_i)$
        \STATE $\mathbf{x}_i^{(0)}
        \leftarrow
        \mathbf{x}_i$
    \ENDFOR
    \STATE Initialize
$\{\mathbf{s}_k^{(0)},\mathbf{u}_k^{(0)}\}_{k=1}^{K}$
    \FOR{$\ell=1$ to $L$}
        \FORALL{$i=1,\ldots,N^t$}
            \STATE $\mathbf{q}_i^{(\ell)}
            \leftarrow
            \mathbf{W}_q\mathbf{x}_i^{(\ell-1)}$
            \STATE $\mathbf{v}_i^{(\ell)}
            \leftarrow
            \mathbf{W}_v\mathbf{x}_i^{(\ell-1)}$
        \ENDFOR
        \STATE $\mathbf{A}^{(\ell)}
        \leftarrow
        \mathrm{SparseAssign}
        \bigl(\{\mathbf{q}_i^{(\ell)},\mathbf{d}_i\},
        \{\mathbf{s}_k^{(\ell-1)},\mathbf{u}_k^{(\ell-1)}\}\bigr)$
        \FORALL{$k=1,\ldots,K$}
            \STATE $\mathbf{e}_k^{(\ell)}
\leftarrow
\mathrm{SemanticAggregate}
(\{a_{ik}^{(\ell)},g_i,\mathbf{v}_i^{(\ell)}\}_{i=1}^{N^t})$
            \STATE $\mathbf{c}_k^{(\ell)}
\leftarrow
\mathrm{DescriptorAggregate}
(\{a_{ik}^{(\ell)},\mathbf{d}_i\}_{i=1}^{N^t})$
            \STATE $\mathbf{s}_k^{(\ell)}
            \leftarrow
            \mathrm{GRU}_{s}
            (\mathbf{s}_k^{(\ell-1)},\mathbf{e}_k^{(\ell)})$
            \STATE $\mathbf{u}_k^{(\ell)}
            \leftarrow
            \mathrm{GRU}_{u}
            (\mathbf{u}_k^{(\ell-1)},\mathbf{c}_k^{(\ell)})$
        \ENDFOR
        \FORALL{$i=1,\ldots,N^t$}
            \STATE $\mathbf{x}_i^{(\ell)}
            \leftarrow
            \mathrm{TokenUpdate}
            (\mathbf{x}_i^{(\ell-1)},\mathbf{A}^{(\ell)},
            \{\mathbf{s}_k^{(\ell)}\})$
        \ENDFOR
    \ENDFOR

    \STATE $\mathbf{o}^t
    \leftarrow
    \mathrm{Pool}
    (\{\mathbf{s}_k^{(L)}\}_{k=1}^{K})$

    \STATE $\hat{\mathbf{y}}^t
    \leftarrow
    \phi(\mathbf{o}^t)$
\ENDFOR

\RETURN $\{\hat{\mathbf{y}}^t\}_{t=1}^{T}$

\end{algorithmic}
\end{algorithm}

\subsection{Hyper-Slot for Multi-dimensional Fusion}
\label{sec:hyperslot}
To fuse heterogeneous semantic tokens received at the edge server, we propose Hyper-Slot, a dimension-aware token fusion module. The key idea is to map tokens with different semantic contents and heterogeneous conditions into a fixed number of latent slots, where each slot adaptively captures a group of tokens that are coherent in both semantic and task-specific dimensions.

We consider a set of semantic tokens
\begin{equation}
\mathcal{X}=\{(\mathbf{x}_i,\eta_i)\}_{i=1}^{N},
\end{equation}
where $\mathbf{x}_i\in\mathbb{R}^{d_x}$ denotes the semantic embedding of token $i$, $N$ is the number of received tokens, and $\eta_i$ denotes its heterogeneity descriptor. Specifically,
\begin{equation}
\eta_i=(\zeta_i^{(1)},\zeta_i^{(2)},\ldots,\zeta_i^{(P)}),
\end{equation}
where $P$ is the number of task-configurable dimensions. Each dimension $\zeta_i^{(p)}$ may describe a categorical or continuous property of the token, such as its source, modality, semantic granularity, confidence, freshness, or communication condition. These dimensions are not fixed by the model and can be instantiated according to the considered edge task.

The heterogeneity descriptor is encoded as
\begin{equation}
\mathbf{d}_i=\Gamma_{\theta}(\eta_i)\in\mathbb{R}^{d_\eta},
\end{equation}
where $\Gamma_{\theta}(\cdot)$ embeds categorical dimensions and maps continuous dimensions into a unified descriptor space.

Hyper-Slot maintains $K$ learnable slots. At iteration $l$, slot $k$ contains a semantic anchor $\mathbf{s}_k^{(l)}\in\mathbb{R}^{d}$ and a heterogeneity anchor $\mathbf{u}_k^{(l)}\in\mathbb{R}^{d_\eta}$. The token state is initialized as $\mathbf{x}_i^{(0)}=\mathbf{x}_i$. At each iteration, token queries and values are computed by
\begin{equation}
\mathbf{q}_i^{(l)}=\mathbf{W}_q\mathbf{x}_i^{(l-1)},\qquad
\mathbf{v}_i^{(l)}=\mathbf{W}_v\mathbf{x}_i^{(l-1)},
\end{equation}
where $\mathbf{W}_q,\mathbf{W}_v\in\mathbb{R}^{d\times d_x}$ are learnable projections.

The affinity between token $i$ and slot $k$ is defined by combining semantic similarity and heterogeneity consistency:
\begin{equation}
\ell_{ik}^{(l)}
=
\frac{\langle \mathbf{s}_k^{(l-1)},\mathbf{q}_i^{(l)}\rangle}{\sqrt{d}}
+
\beta
\frac{\langle \mathbf{u}_k^{(l-1)},\mathbf{d}_i\rangle}{\sqrt{d_\eta}},
\end{equation}
where $\beta$ balances the contribution of the heterogeneity descriptor. The assignment from tokens to slots is obtained by sparse normalization \cite{martins2016softmax}:
\begin{equation}
\label{sparse}
a_{ik}^{(l)}
=
\text{sparsemax}_k
\left(
\ell_{i1}^{(l)},\ldots,\ell_{iK}^{(l)}
\right),
\end{equation}
where $a_{ik}^{(l)}\geq 0$ and $\sum_{k=1}^{K}a_{ik}^{(l)}=1$.

We further introduce a descriptor-dependent gate $g_i$ to dynamically control the effective contribution of each token
\begin{equation}
g_i=\sigma\left(\mathbf{w}_g^{\top}\mathbf{d}_i+b_g\right),
\end{equation}
where $g_i\in(0,1)$ reflects the reliability or usefulness of token $i$ under its heterogeneity descriptor. Each slot aggregates semantic and descriptor information as
\begin{equation}
\mathbf{e}_k^{(l)}
=
\frac{
\sum_{i=1}^{N} a_{ik}^{(l)} g_i \mathbf{v}_i^{(l)}
}{
\sum_{i=1}^{N} a_{ik}^{(l)}+\epsilon
},
\end{equation}
\begin{equation}
\mathbf{c}_k^{(l)}
=
\frac{
\sum_{i=1}^{N} a_{ik}^{(l)} \mathbf{d}_i
}{
\sum_{i=1}^{N} a_{ik}^{(l)}+\epsilon
},
\end{equation}
where $\mathbf{e}_k^{(l)}$ is the aggregated semantic representation, $\mathbf{c}_k^{(l)}$ is the aggregated heterogeneity representation, and $\epsilon$ is a small constant for numerical stability.

The slot anchors are then updated by recurrent refinement based on Gated Recurrent Units (GRUs) \cite{cho2014learning}:
\begin{equation}
\mathbf{s}_k^{(l)}
=
\text{GRU}_s\left(\mathbf{s}_k^{(l-1)},\mathbf{e}_k^{(l)}\right),
\qquad
\mathbf{u}_k^{(l)}
=
\text{GRU}_u\left(\mathbf{u}_k^{(l-1)},\mathbf{c}_k^{(l)}\right).
\end{equation}
To allow tokens to adjust their slot assignments across iterations, the slot information is written back to the token states:
\begin{equation}
\label{LN}
\mathbf{x}_i^{(l)}
=
\text{LN}\left(
\mathbf{x}_i^{(l-1)}
+
\alpha
\sum_{k=1}^{K}
a_{ik}^{(l)}
\psi(\mathbf{s}_k^{(l)})
\right),
\end{equation}
where $\psi:\mathbb{R}^{d}\rightarrow\mathbb{R}^{d_x}$ is a learnable projection, $\LN(\cdot)$ denotes layer normalization \cite{ba2016layer}, and $\alpha$ controls the residual strength.

After $L$ iterations, the fused representation is obtained by pooling the final slots:
\begin{equation}
\mathbf{o}
=
\text{Pool}\left(
\{\mathbf{s}_k^{(L)}\}_{k=1}^{K}
\right),
\end{equation}
where $\Pool(\cdot)$ can be mean pooling or attention pooling. The final prediction is given by
\begin{equation}
\hat{\mathbf{y}}=\phi(\mathbf{o}),
\end{equation}
where $\phi(\cdot)$ is a task-specific prediction head.

The model is trained end-to-end with
\begin{equation}
\mathcal{L}
=
\mathcal{L}_{\mathrm{task}}
+
\lambda_{\mathrm{div}}\mathcal{L}_{\mathrm{div}},
\end{equation}
where $\mathcal{L}_{\mathrm{task}}$ is the downstream task loss. To prevent all slots from collapsing to similar heterogeneity patterns, we use a diversity regularization term over the heterogeneity anchors:
\begin{equation}
\mathcal{L}_{\mathrm{div}}
=
\frac{1}{L}
\sum_{l=1}^{L}
\sum_{k\neq k'}
\left(
\frac{
\langle \mathbf{u}_k^{(l)},\mathbf{u}_{k'}^{(l)}\rangle
}{
\|\mathbf{u}_k^{(l)}\|_2
\|\mathbf{u}_{k'}^{(l)}\|_2
+\epsilon
}
\right)^2,
\end{equation}
where $\lambda_{\mathrm{div}}$ controls the strength of the diversity regularization.

\subsection{Incentive Mechanism for Semantic Uploading}
\label{sec:game}






Section \ref{sec:framework} established the need to coordinate terminal uploading. The edge server aims to obtain sufficient and complementary semantic information for the global fusion task, whereas each terminal incurs computation and communication costs when uploading its local tokens. Therefore, a terminal may not have sufficient incentive to consume additional resources solely for improving the global task.

To coordinate these objectives, we introduce an incentive mechanism in
which the edge server announces a reward pool $\mathcal{R}$ before
semantic uploading. Each terminal then determines its uploading strategy
according to the announced reward and the strategies of the other
terminals. Since the contribution of one terminal depends on the
information provided by the others, the terminal decisions are coupled.
Accordingly, the edge server acts as the leader, while the terminals form
a non-cooperative game under the given reward pool.

\subsubsection{Utility Function of Terminals}

For notational simplicity, the time-slot index $t$ is omitted throughout
this subsection. Under the announced reward pool $\mathcal{R}$, each
terminal determines its uploading strategy by balancing its
contribution-based reward against the computation and communication
energy costs incurred by semantic uploading. The utility of terminal $m$
can hence be expressed as
\begin{equation}
\begin{aligned}
U_m(\mathbf{b}_m;\mathbf{b}_{-m},\mathcal{R})
={}&
\mathcal{R}
\frac{
\Delta_m(\mathbf{b})
}{
\sum_{j\in\mathcal{M}}
\Delta_j(\mathbf{b})
+\epsilon
}
\\
&-
e_{m,\mathrm{cmp}}(\mathbf{b}_m)
-
e_{m,\mathrm{com}}(\mathbf{b}_m),
\end{aligned}
\end{equation}
where
$\mathcal{R}\Delta_m(\mathbf{b})/
(\sum_{j\in\mathcal{M}}\Delta_j(\mathbf{b})+\epsilon)$
is the reward allocated to terminal $m$ from the total reward pool
$\mathcal{R}$, and $\epsilon>0$ is a small constant that keeps the
denominator strictly positive. The terms
$e_{m,\mathrm{cmp}}(\mathbf{b}_m)$ and
$e_{m,\mathrm{com}}(\mathbf{b}_m)$ are its computation and
communication energy costs, respectively. Here,
$\mathbf{b}_m=[b_{m,1},\ldots,b_{m,N_m}]$ is the transmission-quality
strategy of terminal $m$, $\mathbf{b}_{-m}$ denotes the strategies of
the other terminals, and
$\mathbf{b}=(\mathbf{b}_m,\mathbf{b}_{-m})$ is the joint strategy.
Specifically, $b_{m,i}=0$ indicates that token $x_{m,i}$ is not
uploaded, while $b_{m,i}\in(0,1]$ represents its normalized
transmission quality. A larger $b_{m,i}$ indicates that a
higher-quality communication link is assigned to the token.

The fusion contribution of a terminal depends on the joint information
provided by all participating terminals. For a given joint strategy
$\mathbf{b}$, we characterize the contribution of terminal $m$ by the
marginal change in fusion gain resulting from its participation:
\begin{equation}
\Delta_m(\mathbf{b})
=
F(\mathbf{b})
-
F(\mathbf{0}_m,\mathbf{b}_{-m}),
\end{equation}
where $F(\mathbf{b})$ denotes the fusion gain under the joint strategy,
and $F(\mathbf{0}_m,\mathbf{b}_{-m})$ denotes the fusion gain when
terminal $m$ is excluded while the strategies of the other terminals
remain unchanged. Thus, $\Delta_m(\mathbf{b})$ represents the
contribution of terminal $m$ conditioned on the information jointly
provided by the remaining terminals.

The fusion gain consists of information coverage and cross-type
complementarity. The information coverage characterizes the effective
amount of each information type in each slot, while the cross-type
complementarity characterizes the joint contribution of different
information types assigned to the same slot. We hence define the fusion
gain as
\begin{equation}
\begin{aligned}
F(\mathbf{b})
={}&
\sum_{k=1}^{K}
\sum_{p\in\mathcal{P}}
\omega_p
\left(
1-e^{-\lambda z_{k,p}(\mathbf{b})}
\right)
\\
&+
\sum_{k=1}^{K}
\sum_{\substack{p,q\in\mathcal{P}\\p<q}}
\xi_{p,q}
\min
\left\{
1-e^{-\lambda z_{k,p}(\mathbf{b})},
1-e^{-\lambda z_{k,q}(\mathbf{b})}
\right\},
\end{aligned}
\end{equation}
where $\mathcal{P}$ is the set of task-relevant information types,
$\omega_p\geq0$ is the importance weight of type $p$,
$\lambda>0$ is the saturation coefficient, and
$\xi_{p,q}\geq0$ is the complementarity coefficient between
information types $p$ and $q$.

Notably, $1-e^{-\lambda z_{k,p}(\mathbf{b})}$ has a theoretical upper
bound of one and captures the diminishing return caused by increasingly
high transmission quality of repeated or overlapping information in the
same slot. The quantity
\[
\min
\left\{
1-e^{-\lambda z_{k,p}(\mathbf{b})},
1-e^{-\lambda z_{k,q}(\mathbf{b})}
\right\}
\]
characterizes the complementarity between information types $p$ and
$q$, where their joint contribution is limited by the information type
with lower coverage.

The effective amount of type-$p$ information in slot $k$ is defined as
\begin{equation}
z_{k,p}(\mathbf{b})
=
\sum_{m\in\mathcal{M}}
\sum_{i\in\mathcal{I}_{m,p}}
b_{m,i}a_{m,i,k},
\end{equation}
where $\mathcal{I}_{m,p}$ is the index set of type-$p$ tokens generated
by terminal $m$, and $a_{m,i,k}\geq0$ is the Hyper-Slot assignment
weight defined in the previous subsection. Therefore, the effective
contribution of a token depends on its assigned slot and selected
transmission quality.

Terminal $m$ determines its semantic uploading strategy to maximize its
utility under the reward pool and the strategies of the other terminals.
The corresponding optimization problem is

\noindent\textbf{P1:}
\begin{subequations}
\begin{align}
\max_{\mathbf{b}_m}\quad
&
U_m(\mathbf{b}_m;\mathbf{b}_{-m},\mathcal{R}),
\\
\mathrm{s.t.}\quad
&
0\leq b_{m,i}\leq1,
\quad i=1,\ldots,N_m,
\\
&
\sum_{i=1}^{N_m}b_{m,i}
\leq
\bar b_m.
\end{align}
\end{subequations}

The constraint $0\leq b_{m,i}\leq1$ defines the transmission-quality
decision of each token. The value $b_{m,i}=0$ represents no
transmission, while a positive value represents the normalized quality
of the selected communication link. The constraint
$\sum_{i=1}^{N_m}b_{m,i}\leq\bar b_m$ limits the total high-quality
communication resource available to terminal $m$, where $\bar b_m$ is
its total normalized transmission-quality budget. This constraint
prevents all tokens from using the highest-quality links simultaneously
and encourages the terminal to allocate better links to tokens with
larger fusion contributions.

\subsubsection{Utility Function of the Edge Server}

The edge server determines the reward pool by considering the global
fusion gain and the associated incentive and energy costs. Its utility
can hence be expressed as
\begin{equation}
\begin{aligned}
U_{\mathrm{s}}
\left(
\mathcal{R};
\mathbf{b}^{*}(\mathcal{R})
\right)
={}&
F
\left(
\mathbf{b}^{*}(\mathcal{R})
\right)
-
\mathcal{R}
\\
&-
\sum_{m\in\mathcal{M}}
\left[
e_{m,\mathrm{cmp}}
\left(
\mathbf{b}_m^{*}(\mathcal{R})
\right)
+
e_{m,\mathrm{com}}
\left(
\mathbf{b}_m^{*}(\mathcal{R})
\right)
\right].
\end{aligned}
\end{equation}
Here, $F(\mathbf{b}^{*}(\mathcal{R}))$ is the fusion benefit obtained
under the equilibrium terminal strategies, $\mathcal{R}$ is the reward
pool committed by the edge server, and
\[
\sum_{m\in\mathcal{M}}
\left[
e_{m,\mathrm{cmp}}
\left(
\mathbf{b}_m^{*}(\mathcal{R})
\right)
+
e_{m,\mathrm{com}}
\left(
\mathbf{b}_m^{*}(\mathcal{R})
\right)
\right]
\]
denotes the total computation and communication energy cost.

The edge server anticipates the equilibrium response of the terminals
and determines the reward pool by solving

\noindent\textbf{P2:}
\begin{subequations}
\begin{align}
\max_{\mathcal{R}}\quad
&
U_{\mathrm{s}}
\left(
\mathcal{R};
\mathbf{b}^{*}(\mathcal{R})
\right),
\\
\mathrm{s.t.}\quad
&
0\leq\mathcal{R}\leq\mathcal{R}^{\max},
\end{align}
\end{subequations}
where $\mathcal{R}^{\max}$ is the maximum incentive budget available to
the edge server.

For a given reward pool $\mathcal{R}$, the terminals form a
non-cooperative game. The uploading strategy of each terminal affects
the fusion gain and the contribution-based reward allocation through
the joint strategy $\mathbf{b}$. The strategy profile
$\mathbf{b}^{*}(\mathcal{R})$ is a Nash equilibrium if
\begin{equation}
U_m
\left(
\mathbf{b}_m^{*};
\mathbf{b}_{-m}^{*},
\mathcal{R}
\right)
\geq
U_m
\left(
\mathbf{b}_m;
\mathbf{b}_{-m}^{*},
\mathcal{R}
\right),
\end{equation}
for every terminal $m\in\mathcal{M}$ and every feasible strategy
$\mathbf{b}_m$.

\subsubsection{Bi-level Stackelberg--Nash Game Theoretical Algorithm}

The optimization problems for the terminals (P1) and the edge server
(P2) are coupled. The solution of P1 depends on the reward pool
$\mathcal{R}$ announced by the edge server and the transmission-quality
strategies of the other terminals. Meanwhile, the solution of P2
depends on the equilibrium terminal strategies
$\mathbf{b}^{*}(\mathcal{R})$. This coupling forms a bi-level nonlinear
optimization problem. We hence propose a Stackelberg--Nash game
theoretical algorithm to determine the terminal strategies and the
server reward pool. The terminal-side game has a Cournot-type structure
because each terminal selects its transmission-quality quantities while
competing through the shared fusion gain and contribution-based reward
allocation.

For the game analysis, the Hyper-Slot assignment weights
$a_{m,i,k}$ are obtained from the latest Hyper-Slot state and remain
fixed during one execution of the game. They are updated after fusion
and used in the subsequent time slot. Since
$a_{m,i,k}\geq0$, $\omega_p,\xi_{p,q}\geq0$, and
$1-e^{-\lambda z}$ is non-decreasing in $z$,
$F(\mathbf{b})$ is non-decreasing in every $b_{m,i}$. Therefore,
\begin{equation}
\Delta_m(\mathbf{b})
=
F(\mathbf{b})
-
F(\mathbf{0}_m,\mathbf{b}_{-m})
\geq0.
\end{equation}
The constant $\epsilon>0$ ensures that the reward allocation remains
well defined when all terminals provide zero marginal contribution.

The feasible strategy set of terminal $m$ is
\begin{equation}
\mathcal{B}_m
=
\left\{
\mathbf{b}_m
\ \middle|\
0\leq b_{m,i}\leq1,\;
\sum_{i=1}^{N_m}b_{m,i}\leq\bar b_m
\right\}.
\end{equation}
The set $\mathcal{B}_m$ is non-empty, compact, and convex. Given the
strategies of the other terminals, the utility
$U_m(\mathbf{b}_m;\mathbf{b}_{-m},\mathcal{R})$ is continuous over
$\mathcal{B}_m$. Therefore, the best-response problem of each terminal
admits at least one optimal solution.

For a given reward pool $\mathcal{R}$, the terminals simultaneously
determine their transmission-quality strategies. The strategy of one
terminal affects the common fusion gain and the reward shares of all
terminals. Given $\mathbf{b}_{-m}$ and $\mathcal{R}$, the best response
of terminal $m$ is
\begin{equation}
\mathrm{BR}_m
\left(
\mathbf{b}_{-m};\mathcal{R}
\right)
=
\arg\max_{\mathbf{b}_m\in\mathcal{B}_m}
U_m
\left(
\mathbf{b}_m;
\mathbf{b}_{-m},
\mathcal{R}
\right).
\end{equation}

At differentiable points, the marginal utility of assigning a higher
transmission quality to token $x_{m,i}$ is
\begin{equation}
\begin{aligned}
\frac{\partial U_m}{\partial b_{m,i}}
={}&
\mathcal{R}
\frac{
\displaystyle
\frac{\partial\Delta_m}{\partial b_{m,i}}
\left(
\sum_{j\in\mathcal{M}}\Delta_j+\epsilon
\right)
-
\Delta_m
\displaystyle
\sum_{j\in\mathcal{M}}
\frac{\partial\Delta_j}{\partial b_{m,i}}
}{
\left(
\sum_{j\in\mathcal{M}}\Delta_j+\epsilon
\right)^2
}
\\
&-
\frac{\partial e_{m,\mathrm{cmp}}}{\partial b_{m,i}}
-
\frac{\partial e_{m,\mathrm{com}}}{\partial b_{m,i}}.
\end{aligned}
\end{equation}
The quantity
\[
\frac{\partial}{\partial b_{m,i}}
\left[
\mathcal{R}
\frac{\Delta_m(\mathbf{b})}
{\sum_{j\in\mathcal{M}}\Delta_j(\mathbf{b})+\epsilon}
\right]
\]
characterizes the marginal reward obtained by increasing the
transmission quality of token $x_{m,i}$, accounting for the changes in
both terminal $m$'s contribution and the total contribution used for
reward allocation. The derivatives of
$e_{m,\mathrm{cmp}}(\mathbf{b}_m)$ and
$e_{m,\mathrm{com}}(\mathbf{b}_m)$ characterize the corresponding
marginal computation and communication energy costs.

We use damped best-response iteration to seek a Nash equilibrium among the terminals. At iteration $r$, terminal $m$ first solves P1 under the
current strategies of the other terminals and then updates
\begin{equation}
\mathbf{b}_m^{r+1}
=
(1-\eta)\mathbf{b}_m^{r}
+
\eta
\mathrm{BR}_m
\left(
\mathbf{b}_{-m}^{r};\mathcal{R}
\right),
\end{equation}
where $\eta\in(0,1]$ is the damping factor. The procedure terminates when
\begin{equation}
\left\|
\mathbf{b}^{r+1}
-
\mathbf{b}^{r}
\right\|_2
\leq
\delta,
\end{equation}
where $\delta>0$ is the convergence tolerance.

When the procedure converges, its fixed point satisfies
\begin{equation}
\mathbf{b}_m^{*}
\in
\mathrm{BR}_m
\left(
\mathbf{b}_{-m}^{*};\mathcal{R}
\right),
\qquad
\forall m\in\mathcal{M},
\end{equation}
and is taken as the terminal-side Nash equilibrium
$\mathbf{b}^{*}(\mathcal{R})$.

The edge server acts as the leader and anticipates the Nash equilibrium
response $\mathbf{b}^{*}(\mathcal{R})$. It determines the reward pool by
solving
\begin{equation}
\max_{0\leq\mathcal{R}\leq\mathcal{R}^{\max}}
U_{\mathrm{s}}
\left(
\mathcal{R};
\mathbf{b}^{*}(\mathcal{R})
\right).
\end{equation}

We use numerical gradient ascent to search for the reward pool that maximizes the server utility.
The algorithm is initialized with a feasible reward pool
$\mathcal{R}^{0}$, a learning rate $\alpha$, and a convergence
tolerance. At iteration $\ell$, the gradient is approximated by
\begin{equation}
\frac{dU_{\mathrm{s}}}{d\mathcal{R}}
\approx
\frac{
U_{\mathrm{s}}
\left(
\mathcal{R}^{\ell}+h;
\mathbf{b}^{*}(\mathcal{R}^{\ell}+h)
\right)
-
U_{\mathrm{s}}
\left(
\mathcal{R}^{\ell};
\mathbf{b}^{*}(\mathcal{R}^{\ell})
\right)
}{
h
},
\end{equation}
where $h>0$ is a small step size. Each evaluation of the edge server
utility requires solving the terminal-side game to obtain the
corresponding Nash equilibrium.

The reward pool is then updated as
\begin{equation}
\mathcal{R}^{\ell+1}
=
\Pi_{[0,\mathcal{R}^{\max}]}
\left(
\mathcal{R}^{\ell}
+
\alpha
\frac{dU_{\mathrm{s}}}{d\mathcal{R}}
\left(
\mathcal{R}^{\ell}
\right)
\right),
\end{equation}
where $\Pi_{[0,\mathcal{R}^{\max}]}(\cdot)$ projects the updated reward
pool onto its feasible interval.

The outer iteration terminates when
\begin{equation}
\left|
\mathcal{R}^{\ell+1}
-
\mathcal{R}^{\ell}
\right|
\leq
\delta.
\end{equation}
The resulting pair
\begin{equation}
\left\{
\mathbf{b}^{*}(\mathcal{R}^{*}),
\mathcal{R}^{*}
\right\}
\end{equation}
is taken as an approximate Stackelberg--Nash solution of the bi-level
game. The individual reward of each terminal is then obtained directly
from the contribution-based reward allocation rule in $U_m$.

\section{Simulations}
In this section, we evaluate the proposed framework through semantic fusion experiments and game-theoretic simulations. We compare Hyper-Slot with other fusion methods and examine the effects of heterogeneity dimensions, transmission budgets, and its structural
components. We further evaluate Hyper-Slot on cross-modal retrieval and investigate the performance of the semantic uploading mechanism.

\subsection{Simulation Settings}
We evaluate Hyper-Slot on a tri-modal extension of COCO2014 \cite{lin2014microsoft}, comprising 30,000 training pairs and 5,000 validation pairs across 80 categories. Each image is paired with its original textual captions and, following SpokenCOCO \cite{hsu2021text}, a corresponding spoken caption, yielding three modality sources — vision, text, and audio — per sample. This construction provides a natural source of modality heterogeneity in addition to the semantic granularity, arrival time, and channel condition dimensions considered below.

Visual tokens are extracted with a Faster R-CNN (ResNet-101) backbone \cite{anderson2018bottom}, producing region-level features of dimension 2048. Textual tokens are extracted with BERT-base-uncased \cite{devlin2019bert}, producing token-level features of dimension 768. Audio tokens are obtained by first applying the Montreal Forced Aligner (MFA) \cite{mcauliffe2017montreal} to determine word-level time boundaries in the spoken caption, then extracting a 768-dimensional feature per aligned segment with a frozen pre-trained HuBERT-base model \cite{hsu2021hubert}. All raw features are projected through linear layers into a unified embedding dimension $d_x$.

The number of latent slots is set to $K = 16$ and the number of refinement iterations to $L = 3$. The token embedding dimension is $d_x = 256$, the slot dimension is $d = 256$, and the heterogeneity descriptor dimension is $d_\eta = 256$. The affinity balancing coefficient $\beta$ in (20) is set to $1.0$, and the residual strength $\alpha$ in (26) is set to $1.0$. Hyper-Slot is trained end-to-end using the AdamW optimizer with a learning rate of $1\times10^{-4}$, a weight decay of $1\times10^{-4}$, and a batch size of $64$.

For each token, the descriptor $\eta_i$ is instantiated with four dimensions: (i) modality (vision/text/audio), (ii) semantic granularity, and (iii) arrival time and (iv) channel quality, the latter two simulated to emulate asynchronous terminal reporting and time-varying wireless conditions.

All experiments were conducted on the Lancaster University HEC GPU cluster using NVIDIA V100 (32GB) and L40 (48GB) nodes via the gpu-short/gpu-medium queues.







\subsection{Performance Comparison with Other Fusion Methods}
\label{sec:exp2}

We compare Hyper-Slot with five representative fusion methods for
heterogeneous semantic tokens. Concat + MLP~\cite{zhao2019affective}
flattens and concatenates all tokens before applying an MLP predictor. It
does not explicitly preserve token structure or model relations among
tokens. Mean Pool~\cite{chen2021learning} averages all tokens into a single
global representation, and therefore loses token-level differences.
SCAN~\cite{lee2018stacked} uses stacked cross-attention followed by
pooling to model pairwise interactions, but it does not explicitly group
related tokens. Slot Attention~\cite{locatello2020object} performs
competitive slot-based grouping. However, it does not use
heterogeneity-aware affinity or descriptor-dependent gating. Static
HGNN~\cite{feng2019hypergraph} constructs hyperedges according to
predefined rules, such as grouping tokens with the same modality or
semantic granularity. These fixed hyperedges may not capture relations
that span multiple heterogeneity dimensions. In contrast, Hyper-Slot assigns tokens to slots according to both semantic
similarity and heterogeneity consistency. The assignments are refined
through iterative slot-to-token feedback. This allows the model to learn
cross-dimensional token relations without relying on predefined
grouping rules.

Table~\ref{tab:exp2_baseline_comparison} reports overall mAP over five
random seeds together with parameter count, GFLOPs, inference time, and
memory footprint. Hyper-Slot achieves the highest mean mAP
($0.7610 \pm 0.0239$), outperforming the strongest baseline (SCAN and Slot
Attention, both at $0.724$) by over three points. This
indicates that jointly modeling semantic content and heterogeneity
attributes yields both higher accuracy and more stable convergence than
methods. The efficiency columns show that this
gain comes at a moderate computational cost: Hyper-Slot uses more
parameters and GFLOPs than Slot Attention or Static HGNN, reflecting the
additional heterogeneity encoder $\Gamma_\theta(\cdot)$ and gating
mechanism, but remains substantially lighter than Concat + MLP while
achieving markedly higher accuracy.

\begin{figure}[t]
    \centering
        \centering
        \includegraphics[width=0.5\textwidth]{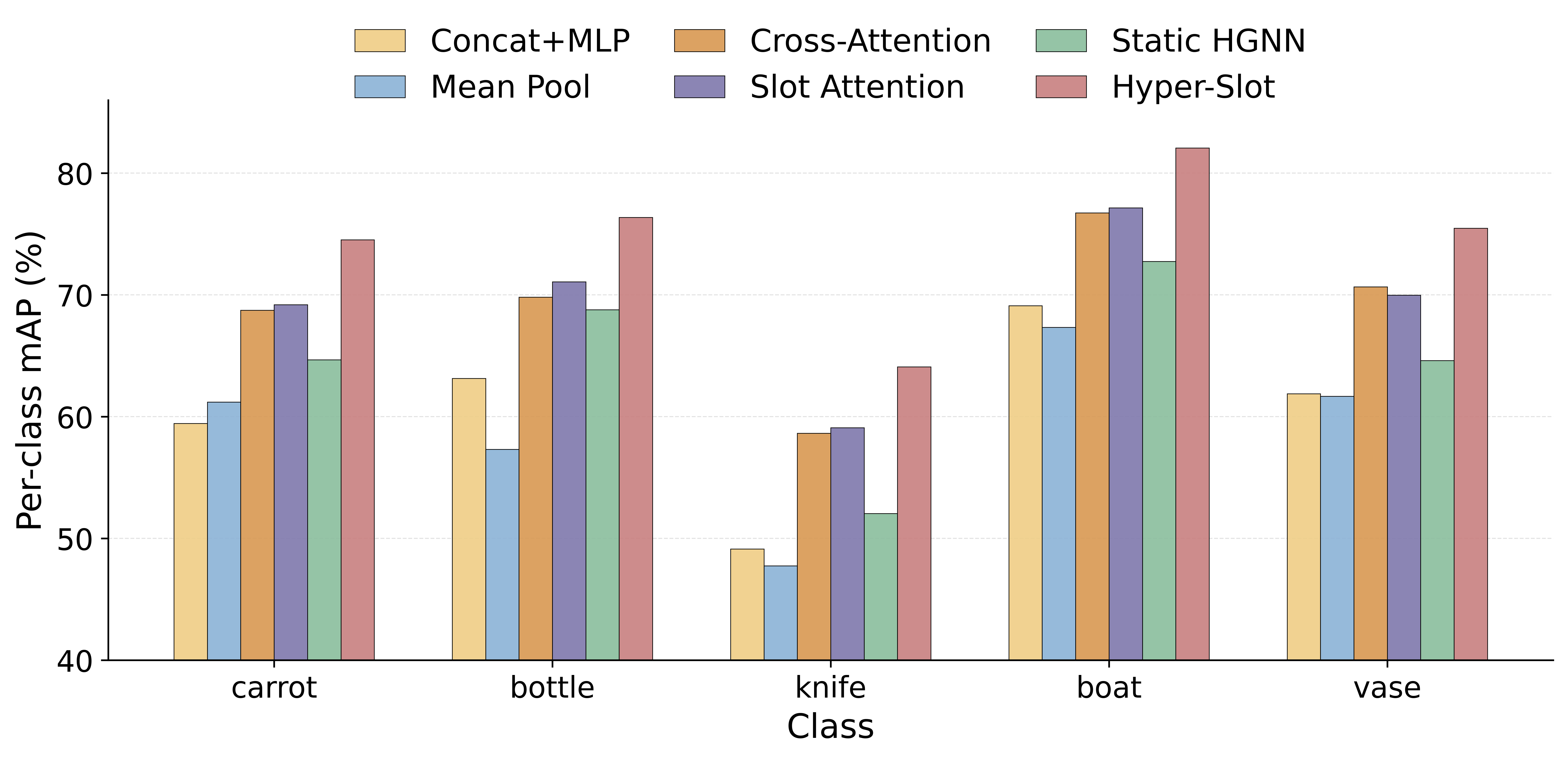}
    \caption{Comparison of per-class mAP across five  categories on COCO2014 dataset.}
    \label{fig:fusion}
    \centering
\end{figure}

\begin{figure}[t]
    \centering
        \centering
        \includegraphics[width=0.5\textwidth]{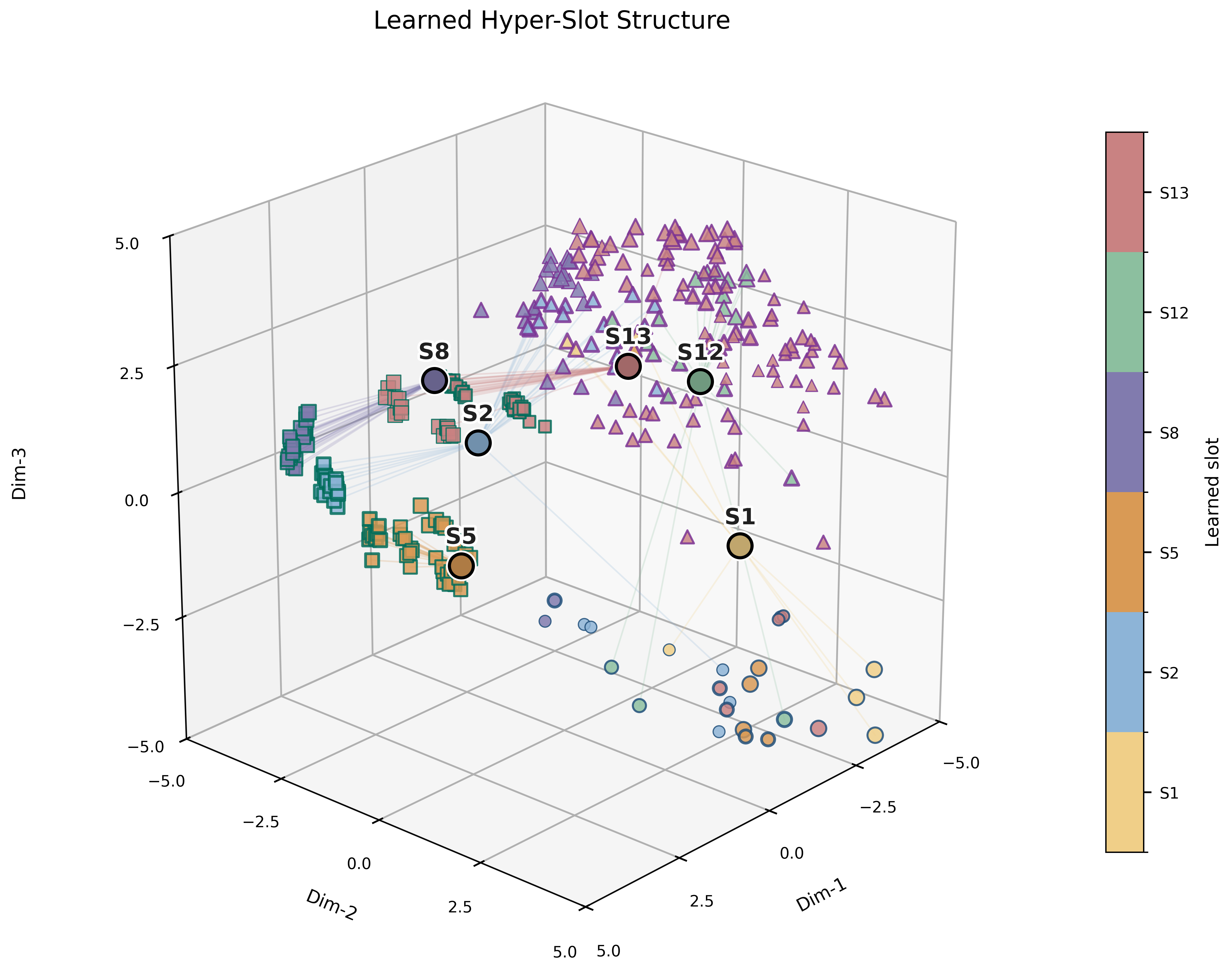}
    \caption{Visualization of token-to-slot assignments in Hyper-Slot. Colors indicate the assigned slots, and star markers denote the slot anchors.}
    \label{fig:visualization}
    \centering
\end{figure}

Figure~\ref{fig:fusion} breaks down per-class mAP across 5 representative
categories. Hyper-Slot achieves the highest mAP in every category shown,
with the margin over the best-performing baseline remaining consistent
across categories of varying difficulty rather than concentrated on a
subset of classes. Figure~\ref{fig:visualization} visualizes the assignment of input tokens to
the learned slots. Tokens of the same modality are distributed across
multiple slots rather than collapsing into one, and each slot anchor lies
close to its assigned token cluster.

\begin{table*}[t]
\centering
\caption{Comparison of mAP and computational efficiency over five random seeds.}
\label{tab:exp2_baseline_comparison}
\resizebox{\textwidth}{!}{
\begin{tabular}{lccccccrrrr}
\toprule
\multirow{2}{*}{Method}
& \multicolumn{6}{c}{mAP over random seeds}
& \multicolumn{4}{c}{Efficiency} \\
\cmidrule(lr){2-7}
\cmidrule(lr){8-11}
& Seed 42 & Seed 123 & Seed 456 & Seed 789 & Seed 1024 & Mean $\pm$ Std
& Params (M) & GFLOPs & Time (ms) & Memory (MB) \\
\midrule
Mean Pool \cite{chen2021learning}
& 0.5431 & 0.5349 & 0.6876 & 0.6964 & 0.6920
& 0.6308 $\pm$ 0.0839
& 1.193 & 0.0582 & 0.0300 & 132.62 \\

Concat MLP \cite{zhao2019affective} 
& 0.6345 & 0.6309 & 0.6386 & 0.6536 & 0.6568
& 0.6429 $\pm$ 0.0116
& 22.034 & 0.0999 & 0.0274 & 450.40 \\

SCAN~\cite{lee2018stacked}
& 0.6564 & 0.6706 & 0.7599 & 0.7702 & 0.7651
& 0.7244 $\pm$ 0.0560
& 3.365 & 0.3364 & 0.0880 & 165.53 \\

Slot Attention \cite{locatello2020object}
& 0.7009 & 0.6969 & 0.7312 & 0.7502 & 0.7407
& 0.7240 $\pm$ 0.0239
& 1.860 & 0.1526 & 0.0565 & 143.38 \\

HGNN \cite{feng2019hypergraph}
& 0.6548 & 0.6469 & 0.7021 & 0.7195 & 0.7108
& 0.6868 $\pm$ 0.0336
& 2.183 & 0.2263 & 0.0720 & 147.94 \\

Hyper-Slot
& 0.7368 & 0.7345 & 0.7724 & 0.7871 & 0.7744
& \textbf{0.76104 $\pm$ 0.0239}
& 4.166 & 0.5729 & 0.1473 & 348.18 \\
\bottomrule
\end{tabular}
}
\vspace{2pt}

\end{table*}



\subsection{Effect of Heterogeneity Dimensions}
\label{sec:exp3}

\begin{figure}[t]
    \centering
        \centering
        \includegraphics[width=0.4\textwidth]{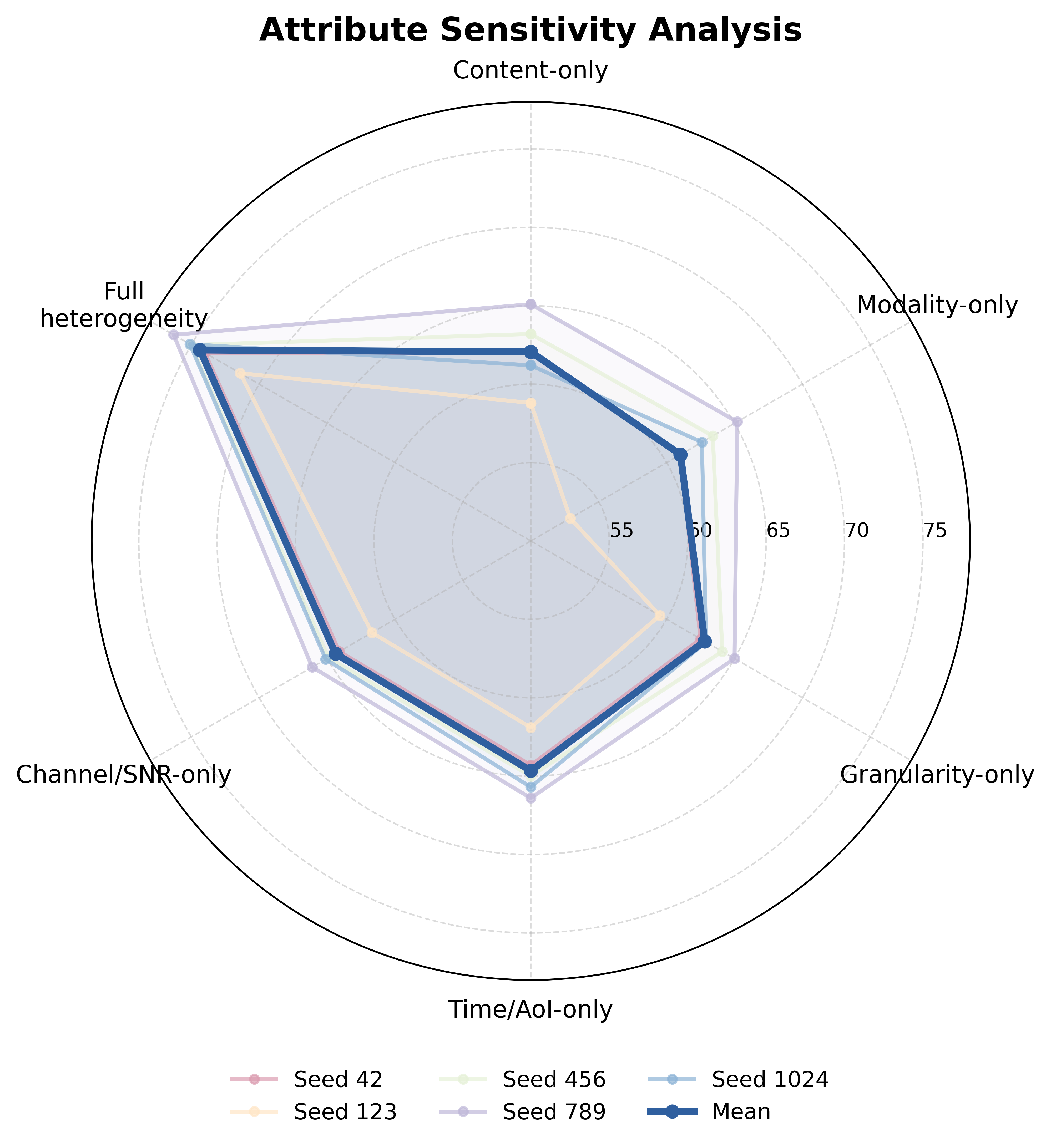}
    \caption{Effect of heterogeneity descriptor configurations on mAP.}
    \label{fig:attribut}
    \centering
\end{figure}

To examine whether the heterogeneity dimensions in $\eta_i$ contribute
independently or jointly shape token contribution, we evaluate six
descriptor configurations: using semantic content alone (i.e., removing the
heterogeneity descriptor entirely), using only modality, only semantic
granularity, only arrival time, only channel quality, and the full
four-dimensional descriptor. Figure~\ref{fig:attribut} reports mAP under
each configuration across five random seeds.

The full heterogeneity descriptor achieves the highest mean mAP across all
seeds, exceeding every single-dimension configuration by a substantial
margin, while content-only, which discards the heterogeneity descriptor
entirely, yields the lowest mean mAP among the six configurations.  The gap between the strongest
single-dimension configuration and the full descriptor remains substantial,
indicating that these dimensions are complementary rather than redundant:
no single attribute recovers the performance achieved by combining all
four. In addition, seed-to-seed variance is markedly larger under
single-dimension configurations than under full heterogeneity, this suggests that relying on
a single heterogeneity attribute not only limits average performance but
also makes fusion quality more sensitive to random initialization. These
results indicate that the heterogeneity dimensions in Hyper-Slot are not
independently informative but interact to shape a token's contribution to
fusion.

\subsection{Effect of Transmission Budget}
\label{sec:exp4}

Terminals operating under limited transmission budgets cannot upload all
candidate tokens and must select a subset to transmit. We compare four
selection strategies against a full-transmission upper bound: Random
selection, which uploads tokens without regard to content; Top-SNR
selection, which prioritizes tokens under the best channel conditions
irrespective of semantic value; Top-importance selection, which ranks
tokens by a task-relevance score; and task-aware selection, which ranks tokens using signals produced by Hyper-Slot: the descriptor-dependent gate $g_i$ in (22) as a proxy for token reliability, and the assignment weight $a_{i,k}$ as a proxy for the marginal coverage a token still contributes to its slot.

\begin{figure}[t]
    \centering
        \centering
        \includegraphics[width=0.4\textwidth]{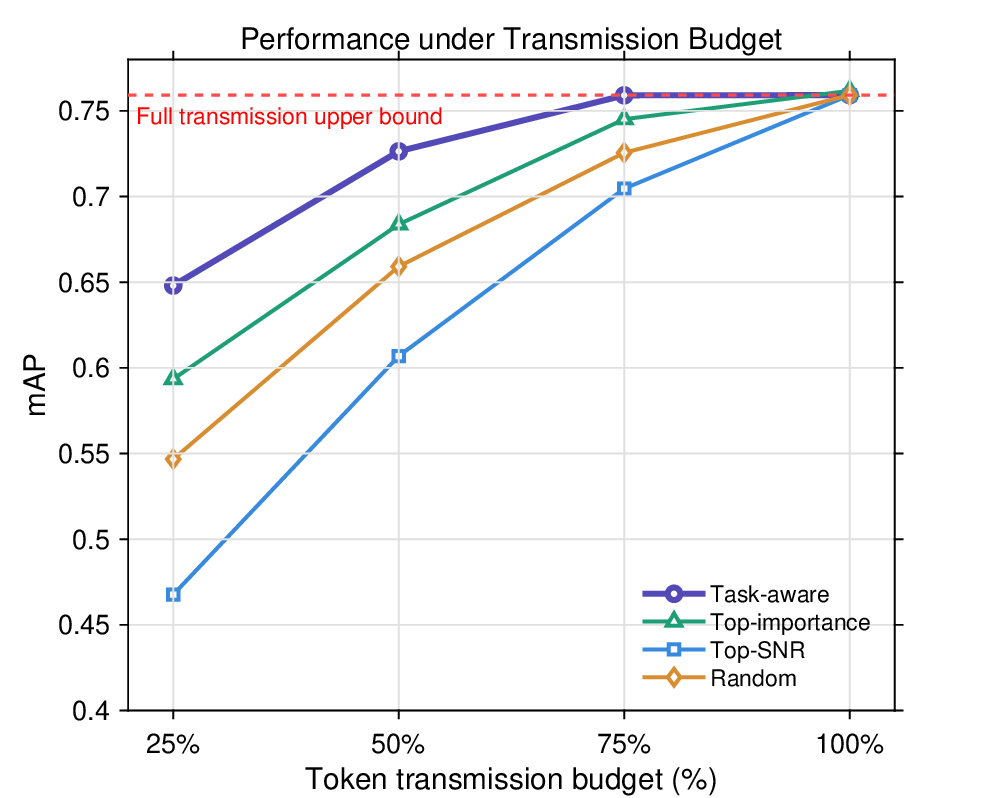}
    \caption{Comparison of token selection strategies under different token transmission budgets. The dashed line indicates the full-transmission performance.}
    \label{fig:exp4}
    \centering
\end{figure}

Figure~\ref{fig:exp4} reports mAP at budgets of 25\%, 50\%, 75\%, and 100\% of the full-transmission cost. Task-aware selection dominates at every budget below 100\%, with the largest margin over the next-best strategy (Top-importance) at 25\% and a narrowing gap as the budget increases, converging to the full-transmission upper bound at 100\%. Top-importance consistently outperforms Random and Top-SNR below 100\% budget. These results show that selection guided by Hyper-Slot's own reliability and coverage signals is more communication-efficient than content-agnostic (Random) or channel-only (Top-SNR) heuristics.

\subsection{Structural Ablation of Hyper-Slot}
\label{sec:exp5}

Hyper-Slot has two core design elements: dependency modeling among tokens
through iterative competitive assignment and slot-to-token feedback, and
the compact fixed-size slot representation. We isolate their contributions
by comparing two structural variants against the full model. Dependency
only retains the iterative sparsemax assignment in~\eqref{sparse} and the
slot-to-token feedback in~\eqref{LN}. It removes the compact bottleneck by
allowing the number of slots to scale with the number of received tokens
rather than remaining fixed at $K$. Compact only retains the fixed
$K$-slot bottleneck. It replaces the iterative competitive assignment and
feedback with a single-pass, non-competitive aggregation.

\begin{figure}[t]
    \centering
        \centering
        \includegraphics[width=0.4\textwidth]{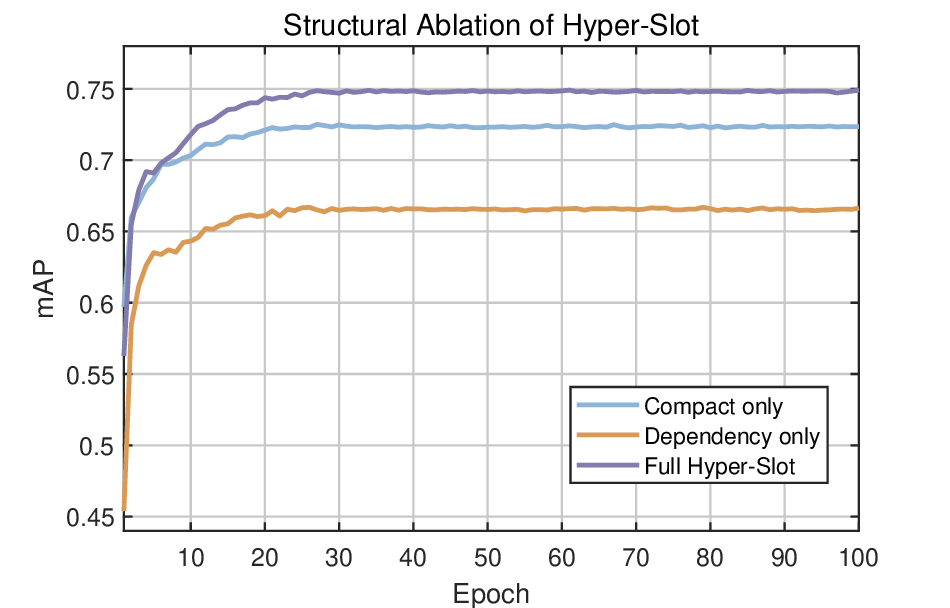}
    \caption{Structural ablation of Hyper-Slot: mAP over training epochs for the compact-only variant, the dependency-only variant, and the full model.}
    \label{fig:exp5}
    \centering
\end{figure}

Figure~\ref{fig:exp5} shows mAP over training epochs for both variants and
the full Hyper-Slot. Full Hyper-Slot converges to the highest mAP
($\approx 0.75$). Compact only converges next ($\approx 0.72$). Dependency
only converges to the lowest mAP ($\approx 0.665$). All three variants
converge within the first 20--30 epochs. Removing the compact bottleneck
causes the larger performance drop. Removing iterative dependency modeling
causes a smaller but consistent drop. Neither variant recovers the mAP of
the full model. This shows the two design elements are complementary.

\subsection{Performance Evaluation on Flickr30K}
\label{sec:flickr}

To verify the applicability of our proposed method across different datasets and tasks, we extended the proposed Hyper-Slot to a dataset for the task of bimodal image-text matching and conducted an additional experiment. Flickr30K~\cite{young2014image} contains 31,000 images with five
captions per image. We follow the standard split of 29,000 training
images, 1,000 validation images, and 1,000 test images. Each image
forms five ground-truth image-caption pairs.

The task is cross-modal retrieval rather than multi-label classification.
Given a query image, the model retrieves the matching caption from the
candidate set, and given a query caption, the model retrieves the matching
image. Performance is measured by RSum, the sum of R@1, R@5, and R@10 for
both image-to-text and text-to-image retrieval.

We compare Hyper-Slot against six baselines.

The concatenation-based fusion baseline~\cite{chen2021learning}, Slot Attention ~\cite{locatello2020object}, and the
fixed-hyperedge baseline following HGNN~\cite{feng2019hypergraph} are the same methods introduced
in Section~V-B, here applied to region and caption token features instead of
the tri-modal tokens used earlier.
VSE++~\cite{faghri2017vse++}
aggregates region features by simple pooling before matching. 
GPO~\cite{chen2021learning} learns a generalized pooling function over
region features. The fixed-hyperedge baseline follows
HREM~\cite{fu2023learning}
models relations among fragments and instances through self-attention.

Table~\ref{tab:flickr_baseline_rsum} reports RSum over five random seeds
together with parameter count, GFLOPs, inference time, and memory usage.
Hyper-Slot achieves the highest mean RSum ($478.9 \pm 3.4$), exceeding the
strongest baseline, Slot Attention ($463.0 \pm 3.2$), by 15.9 points.
HREM is the second-strongest baseline ($459.2 \pm 3.2$). GPO and the fixed
hyperedge baseline follow, both above VSE++ and the concatenation baseline.
VSE++ has the fewest parameters among all methods but the lowest RSum. Hyper-Slot uses fewer parameters than
the concatenation baseline and Slot Attention, while achieving the highest
accuracy among all seven methods.

\begin{table*}[t]
\centering
\caption{Comparison of RSum and computational efficiency on Flickr30K over five random seeds.}
\label{tab:flickr_baseline_rsum}
\resizebox{\textwidth}{!}{
\begin{tabular}{lccccccrrrr}
\toprule
\multirow{2}{*}{Method}
& \multicolumn{6}{c}{RSum over random seeds}
& \multicolumn{4}{c}{Efficiency} \\
\cmidrule(lr){2-7}
\cmidrule(lr){8-11}
& Seed 42 & Seed 123 & Seed 456 & Seed 789 & Seed 1024 & Mean $\pm$ Std
& Params (M) & GFLOPs & Time (ms) & Memory (MB) \\
\midrule
VSE++~\cite{faghri2017vse++}
& 411.2 & 417.5 & 412.9 & 419.0 & 415.1
& 415.1 $\pm$ 3.2
& 2.886 & 0.1825 & 0.0330 & 31.39 \\

Concat MLP \cite{chen2021learning}
& 428.6 & 433.1 & 430.5 & 436.2 & 432.4
& 432.2 $\pm$ 2.9
& 80.484 & 0.3377 & 0.0632 & 330.15 \\

GPO \cite{chen2021learning} 
& 446.8 & 451.5 & 454.2 & 456.0 & 450.9
& 451.9 $\pm$ 3.5
& 2.911 & 0.1826 & 0.1267 & 50.52 \\

HGNN \cite{feng2019hypergraph}
& 436.4 & 443.1 & 439.8 & 446.2 & 441.5
& 441.4 $\pm$ 3.7
& 5.018 & 0.3056 & 0.1025 & 49.13 \\

Slot Attention \cite{locatello2020object}
& 458.7 & 463.4 & 461.2 & 467.0 & 464.8
& 463.0 $\pm$ 3.2
& 26.006 & 1.3289 & 0.3087 & 132.91 \\

HREM~\cite{fu2023learning}
& 454.9 & 460.3 & 457.6 & 463.5 & 459.8
& 459.2 $\pm$ 3.2
& 15.489 & 0.7697 & 0.2210 & 93.03 \\

Hyper-Slot
& 474.6 & 479.2 & 476.8 & 483.1 & 481.0
& \textbf{478.9 $\pm$ 3.4}
& 21.867 & 1.3553 & 0.7147 & 117.56 \\
\bottomrule
\end{tabular}
}
\vspace{2pt}
\begin{flushleft}
\footnotesize

\end{flushleft}
\end{table*}

\subsection{Performance of the Incentive Uploading Mechanism}

To validate the efficacy of the proposed incentive mechanism, we conduct a game-theoretic simulation. This experiment is designed to show the optimal incentive strategy for semantic uploading by analysing the relationship between the server utility and the reward pool. We configure a network with five terminals, three information types, and eight fusion slots. The number of semantic tokens generated by each terminal is randomly varied from 8 to 16 to emulate heterogeneous information availability. The information-type importance weights are set to $\boldsymbol{\omega}=[1.0,1.0,1.2]$, while the cross-type complementarity coefficients $\xi_{p,q}$ are set between 0.30 and 0.40. The saturation coefficient is set to $\lambda=1$. For each terminal, the transmission-quality budget is randomly set between 50\% and 60\% of its full-quality transmission demand. The normalized computation and communication cost coefficients are randomly varied within $[0.05,0.10]$ and $[0.10,0.20]$, respectively, to emulate heterogeneous resource costs. The maximum reward pool is set to $\mathcal{R}_{\max}=12$. For numerical solution, the damping factor of the terminal-side best-response iteration is set to $\eta=0.3$, with a convergence tolerance of $10^{-4}$, while the finite-difference step for the server-side optimization is set to $h=10^{-2}$.

\begin{figure}[h]
    \centering
        \includegraphics[width=0.48\textwidth]{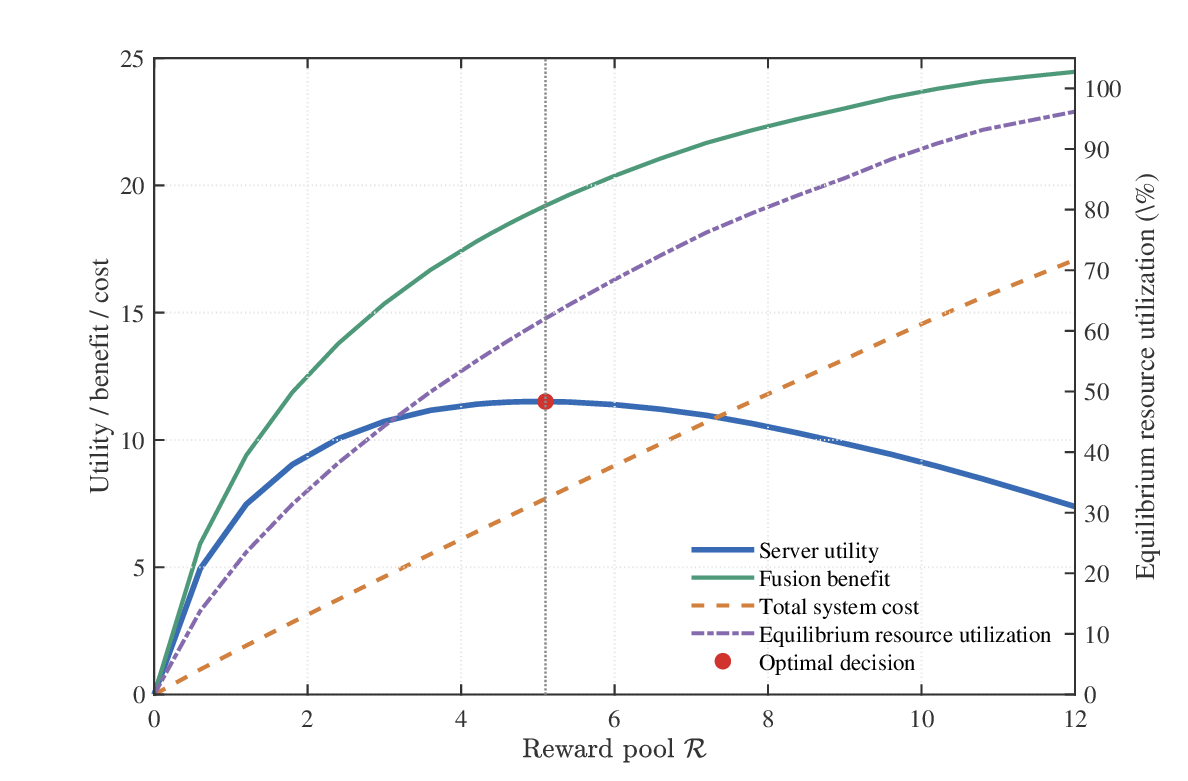}
    \caption{Effect of the reward pool on server utility, fusion benefit, total system cost, and equilibrium resource utilization.}
    \label{fig:game1}
\end{figure}

Fig.~\ref{fig:game1} shows the equilibrium results under different reward pools. As $\mathcal{R}$ increases, the equilibrium resource utilization gradually increases, indicating that stronger incentives encourage terminals to allocate more transmission resources to semantic uploading. This increased resource allocation improves the semantic fusion benefit, although the gain becomes progressively smaller due to the saturation effect. At the same time, the total system cost continuously increases because of the larger reward expenditure and the higher resource consumption of the terminals. Consequently, the server utility $U_{\mathrm{s}}(\mathcal{R})$ first increases and then decreases, reaching its maximum at approximately $\mathcal{R}^{*}=4.91$. When the reward pool is small, the incentive is insufficient to induce adequate resource allocation, resulting in limited fusion gain. When the reward pool becomes excessively large, the additional fusion benefit is no longer sufficient to compensate for the increased system cost. The optimal equilibrium therefore represents a balance among the incentive level, resource utilization, semantic fusion benefit, and system cost.

\begin{figure}[!htbp]
\centering
\subfloat[{Semantic contribution}]{
\begin{minipage}[b]{0.25\textwidth}
    \includegraphics[width=1.8in]{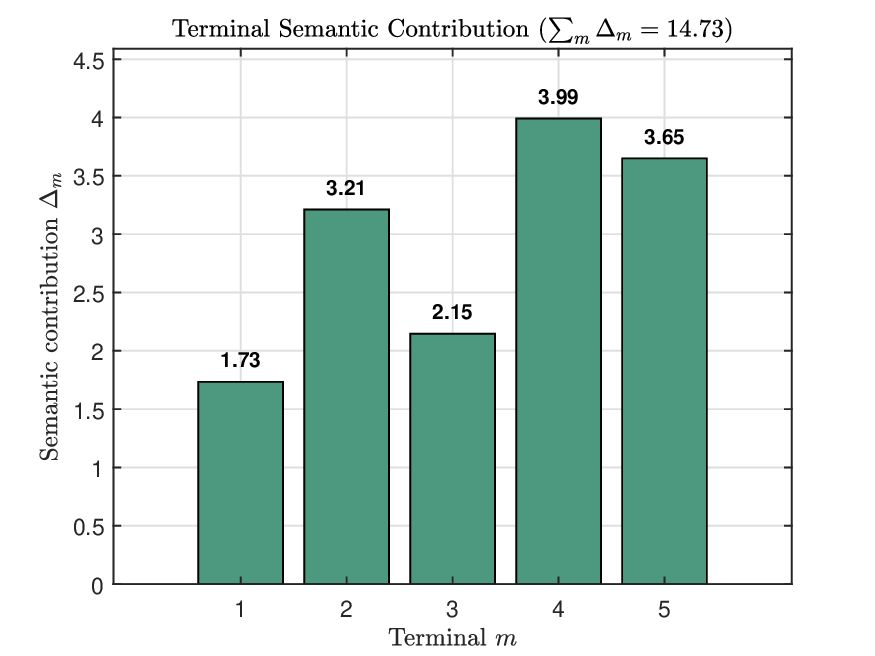}
\end{minipage}}
\subfloat[{Allocated reward}]{
\begin{minipage}[b]{0.25\textwidth}
    \includegraphics[width=1.8in]{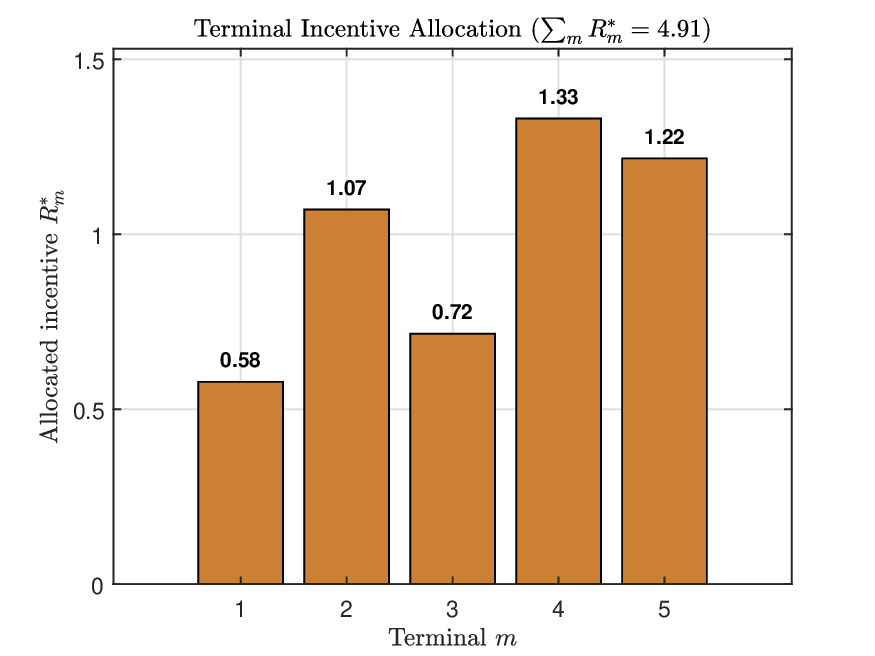}
\end{minipage}}
\caption{{Terminal contributions and reward allocation under the uploading strategies}}
\label{fig:game2}
\end{figure}

Fig.~\ref{fig:game2} (a) shows the semantic contribution of each terminal under the optimal strategy. The contributions differ noticeably among the five terminals, reflecting their heterogeneous semantic information, token assignments, and resource conditions. Terminal 4 provides the largest semantic contribution, while Terminal 1 provides the smallest. This differentiated contribution indicates that the value of a terminal is determined by the additional information it provides to the global fusion task rather than simply by the amount of information uploaded. Fig.~\ref{fig:game2} (b) further shows the resulting incentive allocated to each terminal. The incentive distribution follows the semantic contributions, with Terminal 4 receiving the largest incentive and Terminal 1 receiving the smallest. Meanwhile, the total allocated incentive satisfies $\sum_{m\in\mathcal{M}}R_m^{*}\approx\mathcal{R}^{*}$. The variations in individual incentives therefore reflect the different contributions of heterogeneous terminals to the global fusion task. This resulting differentiated allocation validates the ability of the proposed game mechanism to establish a contribution-based incentive strategy for semantic uploading under heterogeneous terminal conditions.

\section{Conclusion}


In this paper, we proposed a multi-source and multi-dimensional semantic information fusion framework for resource and communication constrained edge networks. The framework coordinates selective semantic uploading among multiple terminals and fuses the resulting incomplete and heterogeneous tokens at the edge server into a compact, task-level representation, realized through a novel Hyper-Slot module for multi-dimensional token
fusion together with a semantic uploading mechanism based on a bi-level Stackelberg--Nash game. With Hyper-Slot, tokens are assigned to latent slots according to both semantic content and multi-dimensional attributes through iterative token-to-slot assignment and slot-to-token feedback, allowing incomplete and variable-size token sets to be aggregated into a robust representation for global inference. The uploading mechanism evaluates terminal contributions through the Hyper-Slot assignment weights and establishes a contribution-based incentive strategy that aligns individual uploading decisions with global fusion benefit under resource
constraints. Hyper-Slot outperforms the strongest evaluated baseline by 3.66 percentage points in mAP on a tri-modal extension of COCO2014 and by 15.9 points in RSum on Flickr30K, requiring less than one-fifth of the parameters of a naive concatenation-based approach and sustaining a 5.46-percentage-point mAP advantage over token selection strategy under a constrained 25\% transmission budget.

\bibliographystyle{ieeetr}
\bibliography{ref}

\end{document}